\documentclass[aip,rsi,reprint]{revtex4-1}
\usepackage{graphicx}
\usepackage{amsmath, amssymb, amsthm}
\usepackage{url}
\usepackage{appendix}
\usepackage{color}
\usepackage{tikz}
\usetikzlibrary{shapes,arrows}
\usepackage{bm}
\usepackage{subfigure}

\begin{document}

\title{An application of a Si/CdTe Compton camera for the polarization measurement of hard x-rays from highly charged heavy ions}
\author{Yutaka Tsuzuki}
\email{\url{yutaka.tsuzuki@ipmu.jp}}
\affiliation{Department of Physics, The University of Tokyo, 7-3-1, Hongo, Bunkyo, Tokyo 113-0033, Japan}
\affiliation{Kavli Institute for the Physics and Mathematics of the Universe (WPI), Institutes for Advanced Study (UTIAS), The University of Tokyo, 5-1-5 Kashiwa-no-Ha, Kashiwa, Chiba, 277-8583, Japan}
\author{Shin Watanabe}
\affiliation{Institute of Space and Astronautical Science, Japan Aerospace Exploration Agency, 3-1-1 Yoshinodai, Chuo Sagamihara, Kanagawa, 252-5210, Japan}
\affiliation{Kavli Institute for the Physics and Mathematics of the Universe (WPI), Institutes for Advanced Study (UTIAS), The University of Tokyo, 5-1-5 Kashiwa-no-Ha, Kashiwa, Chiba, 277-8583, Japan}
\author{Shimpei Oishi}
\author{Nobuyuki Nakamura}
\affiliation{Institute for Laser Science, The University of Electro-Communications, Chofu, Tokyo 182-8585, Japan}
\author{Naoki Numadate}
\affiliation{Institute for Laser Science, The University of Electro-Communications, Chofu, Tokyo 182-8585, Japan}
\affiliation{Komaba Institute for Science, The University of Tokyo, 3-8-1 Komaba, Meguro, Tokyo 153-8902, Japan}
\author{Hirokazu Odaka}
\affiliation{Department of Physics, The University of Tokyo, 7-3-1, Hongo, Bunkyo, Tokyo 113-0033, Japan}
\affiliation{Kavli Institute for the Physics and Mathematics of the Universe (WPI), Institutes for Advanced Study (UTIAS), The University of Tokyo, 5-1-5 Kashiwa-no-Ha, Kashiwa, Chiba, 277-8583, Japan}
\author{Yuusuke Uchida}
\affiliation{Department of Physics, Hiroshima University, 1-3-1 Kagamiyama, Higashi-Hiroshima, Hiroshima 739-8526, Japan}
\author{Hiroki Yoneda}
\affiliation{RIKEN Nishina Center, 2-1 Hirosawa, Wako, Saitama 351-0198, Japan}
\author{Tadayuki Takahashi}
\affiliation{Kavli Institute for the Physics and Mathematics of the Universe (WPI), Institutes for Advanced Study (UTIAS), The University of Tokyo, 5-1-5 Kashiwa-no-Ha, Kashiwa, Chiba, 277-8583, Japan}
\affiliation{Department of Physics, The University of Tokyo, 7-3-1, Hongo, Bunkyo, Tokyo 113-0033, Japan}
\date{\today}

\begin{abstract}
The methods to measure the polarization of the x-rays from highly charged heavy ions with a significantly higher accuracy than the existing technology is needed to explore relativistic and quantum electrodynamics (QED) effects including the Breit interaction. We developed the Electron Beam Ion Trap Compton Camera (EBIT-CC), a new Compton polarimeter with pixelated multi-layer silicon and cadmium telluride counters. The EBIT-CC detects the three-dimensional position of Compton scattering and photoelectric absorption, and thus the degree of polarization of incoming x-rays can be evaluated. We attached the EBIT-CC on the Tokyo Electron Beam Ion Trap (Tokyo-EBIT) in the University of Electro-Communications. An experiment was performed to evaluate its polarimetric capability through an observation of radiative recombination x-rays emitted from highly charged krypton ions, which were generated by the Tokyo-EBIT. The Compton camera of the EBIT-CC was calibrated for the $\sim 75 {\rm\ keV}$ x-rays. We developed event reconstruction and selection procedures and applied them to every registered event. As a result, we successfully obtained the polarization degree with an absolute uncertainty of $0.02$. This uncertainty is small enough to probe the difference between the zero-frequency approximation and full-frequency-dependent calculation for the Breit interaction, which is expected for dielectronic recombination x-rays of highly charged heavy ions.
\end{abstract}

\maketitle

\section{Introduction}
\label{sec:intro}

Polarization of hard x-rays emitted through interactions between highly charged heavy ions and electrons is of significant importance for diagnostics of relativistic quantum effects\cite{nakamura} as well as of non-thermal components in hot plasma\cite{haug}. An electron beam ion trap (EBIT) is a powerful device for studying such polarized x-rays, with which one can observe x-rays emitted from trapped highly charged ions interacting with a unidirectional mono-energy electron beam. \par 
Polarization measurements with an EBIT have been performed using mainly two kinds of polarimeters so far. One of them is Bragg crystal polarimeters\cite{bragg-crystal}. They are useful for x-rays with energies $\lesssim 10 {\rm\ keV}$ and thus are used for measuring the polarization of K x-rays from ions with an atomic number $Z \lesssim 30$ or L x-rays from med-$Z$ ($\sim 50$) ions. For example, Beiersdorfer et al.\cite{beiersdorfer} measured the polarization of x-rays with energies $\simeq 6.7 {\rm\ keV}$ from helium-like iron (Fe) using a crystal polarimeter. The other is Compton polarimeters which employ light materials such as beryllium (Be). Their typical energy range is $10 \, - \, 30 {\rm\ keV}$ and hence they are suitable for measuring the polarization of K x-rays emitted through dielectronic recombination (DR) from highly charged krypton (Kr) and xenon (Xe). For example, Shah et al.\cite{shah} measured the degree of polarization of DR x-rays emitted from highly charged Kr ions, using a Compton polarimeter with a Be scatterer and an absorber with Si-PIN diodes. \par 
By contrast, for K x-rays of heavier ions with $Z \gtrsim 80$, where strong relativistic and quantum electrodynamics (QED) effects are expected, a polarimeter sensitive for a higher energy range, such as $70$ - $80 {\rm\ keV}$, is needed. Furthermore, higher sensitivity and polarimetric capability than the existing polarimeters are also necessary to explore the nature of the relativistic and QED effects, such as the limitation of the zero-frequency approximation in the Breit interaction\cite{mann, nakamura, fritzsche, nakamura-evidence}. The Breit interaction is often calculated by assuming the frequency of the virtual photon that is exchanged between the interacting electrons to be zero. This zero-frequency approximation generally yields sufficient accuracy, {\it i.e.}, the difference between the approximation and full-frequency-dependent calculation is small in most cases. Tong et al.\cite{tong} calculated the polarization of x-rays emitted in dielectronic recombination of lithium-like ions and showed that the approximation modifies the degree of polarization by $\sim 0.02$ for heavy elements with $Z \sim 80$. It is still difficult to probe the difference between the zero-frequency approximation and full-frequency-dependent calculation with any of the methods stated above. \par
Recently, novel Compton cameras that employ silicon (Si) and cadmium telluride (CdTe) semiconductor detectors have been developed as promising polarimeters for x-ray and gamma-ray observations in field of astrophysics\cite{takahashi, watanabe-cc}. Indeed, a Si/CdTe Compton camera, developed for observations of high-energy celestial objects, was adopted as the Soft Gamma-ray Detector (SGD) onboard the {\it Hitomi\,} satellite\cite{watanabe}. {\it Hitomi\,} has succeeded in detecting the polarized hard x-ray emission from the Crab nebula\cite{hitomi-collaboration}. \par
Katsuta et al.\cite{katsuta} studied the polarimetric capability of a Si/CdTe Compton camera with experiments using a synchrotron beam facility, SPring-8, and demonstrated that the Si/CdTe Compton camera had excellent polarimetric performance of detecting the degree of polarization with a $\lesssim 3 {\rm\%}$ uncertainty, proving its great potential. \par
We have developed a new Si/CdTe Compton camera containing pixelated multi-layer semiconductor detectors. It is capable of determining the three-dimensional position of interactions between incoming photons and detector materials inside each layer, and therefore the degree of polarization of the photons can be determined with it more precisely than any other methods to date. In addition, large thickness of the stacked layers ($\sim 2 {\rm\ cm}$ for the Si layers) results in high efficiency ($\sim 50 \%$ at around $70 {\rm\ keV}$) for detecting photons. \par
Our question is whether the polarimetry of hard x-rays emitted from highly charged ions advances by applying the Compton camera. Here we demonstrate the polarimetric performance of our Si/CdTe Compton camera by measuring radiative recombination x-rays emitted from highly charged Kr ions that are generated by an EBIT.\par
In this paper we present the principles and methods of the polarimetry with our Compton camera and the results of the experiment. In Section \ref{sec:cc}, we outline the principles and specifications of the Compton camera. In Section \ref{sec:exp}, the experimental setup and conditions are explained. In Section \ref{sec:ana}, we describe the method of the calibration, Monte Carlo simulation, and data analysis. Section \ref{sec:res} gives the result of the experiment and discussions on the systematic uncertainties of our result. Summary and conclusion are presented in Section \ref{sec:sum}.

\section{Compton camera}
\label{sec:cc}

\subsection{Principle}

\begin{figure}[t]
	\centering
	\includegraphics[width=0.8\linewidth,keepaspectratio]{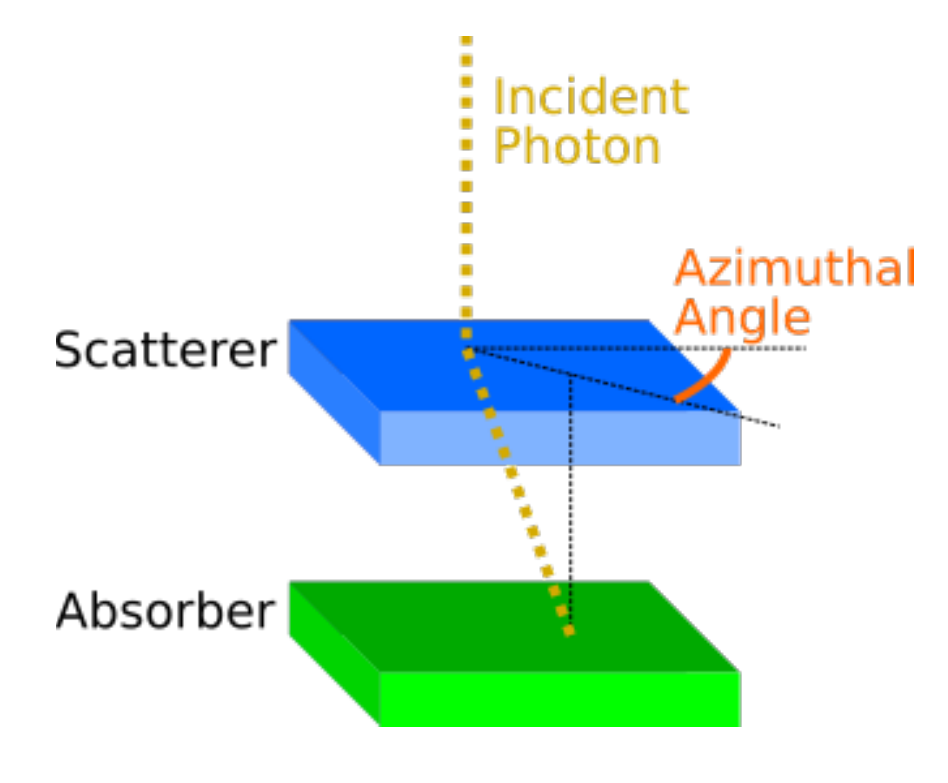}
	\caption{A schematic drawing of polarization measurement with a Compton camera.}
	\label{fig:scheme-cc}
\end{figure}

Compton cameras utilize Compton scattering to obtain information of energy and direction of incoming photons. One can also evaluate the degree of polarization by analyzing the scattering angles of photons. In general, a Compton camera consists of at least one scatterer and one absorber (see Fig. \ref{fig:scheme-cc}). In a Compton camera, incident x-ray photons can be scattered by the scatterer and then are photo-absorbed by the absorber\footnote{Here we simplify the processes of photons. In fact, other sequences of processes are possible. For example, a photon can be scattered by the scatterer and also absorbed by the scatterer.}.
 \par
Since the differential cross section for Compton scattering depends on the polarization angle and is modulated by a squared cosinusoidal term, the degree of polarization of an incident photon beam is measurable with Compton cameras by counting the events where photons are scattered by the scatterer and are absorbed by the absorber. The differential cross section for Compton scattering off a free electron at rest is given by the following Klein-Nishina formula\cite{klein-nishina}:
\begin{eqnarray}
	\frac{d\sigma}{d\Omega} = \frac{{r_0}^2}{2} \left( \frac{E}{E_0} \right)^2 \left( \frac{E}{E_0} + \frac{E_0}{E} - 2 \sin^2 \theta \cos^2 (\phi - \phi_0 ) \right) , \nonumber \\ \label{eq:klein-nishina}
\end{eqnarray}
where $r_0 = 2.818\times10^{-15} {\rm\ m}$ is the classical electron radius, $E_0$ is the initial energy of the incident photon, $E$ is the energy of the scattered photon, $\theta$ is the scattering polar angle, $\phi$ is the azimuthal scattering angle, and $\phi_0$ is the polarization angle of the photon. In reality, electrons are bound to atoms and are not at rest, and thus the distribution of the scattered photons is affected by the so-called Doppler broadening\cite{ordonez,zoglauer}. \par
The degree of polarization of incoming x-rays can be calculated by comparing the two distributions of $\phi$, namely the measured and simulated ones. We usually employ Monte Carlo simulations to obtain the azimuthal distribution model of the events $d^{\rm (model)}(\phi)$ and fit the experimental distribution $d^{\rm (exp)}(\phi)$ with it (see Sec. \ref{sec:res} for detail). By dividing $d^{\rm (exp)}(\phi)$ by the simulated distribution for an unpolarized beam $d^{(0)}(\phi)$, we obtain a modulation curve that contains a component proportional to $\cos ( 2\phi - {\rm const.})$. 

\subsection{The Electron Beam Ion Trap Compton camera (EBIT-CC)}

\begin{figure}[htbp]
    \centering
    \subfigure[]{
        \includegraphics[width=0.8\linewidth,keepaspectratio,page=1]{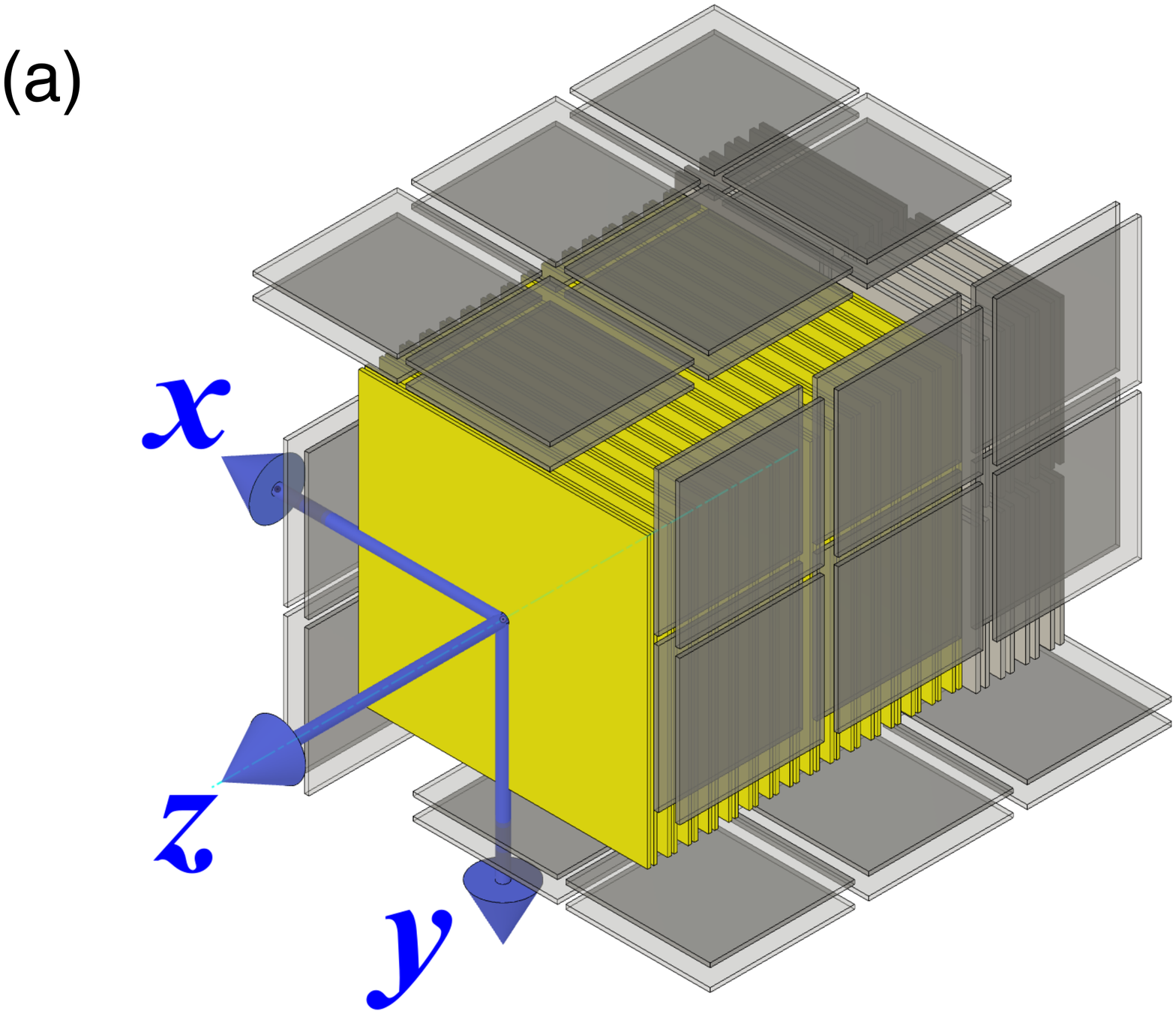}
    } \\
    \subfigure[]{
        \includegraphics[width=0.8\linewidth,keepaspectratio,page=2]{./fig_CC.pdf}
    } \\
    \subfigure[]{
        \includegraphics[width=0.8\linewidth,keepaspectratio,page=3]{./fig_CC.pdf}
    }
    \caption{Schematics and coordinates definition of the detector parts of the EBIT-CC. Yellow and gray parts show the Si and the CdTe detectors, respectively. (a) A schematic of the coordinate system of the EBIT-CC. Its origin is located at the center on the first layer of the Si detectors. (b) Definition of the geometrical scattering angle $\theta_G$. Blue dotted line depicts an incoming x-ray from the source $\vec{x}_{\rm source}$, which is scattered at $\vec{x_1}$ and absorbed at $\vec{x_2}$. (c) Definition of the azimuthal angle $\phi$.}
    \label{fig:sgd-geometry}
\end{figure}
We developed a new type of Si/CdTe Compton camera system for experiments with an EBIT, named the Electron Beam Ion Trap Compton camera (EBIT-CC). It has the same structure as that of the Compton camera of the {\it Hitomi} SGD. We used a new set of parameters of electric devices so that the EBIT-CC covers lower energy range compared to the SGD. The dimensions of the EBIT-CC are roughly $12 {\rm\ cm} \times 12 {\rm\ cm} \times 12 {\rm\ cm}$. \par
The Compton camera in the EBIT-CC employs Si detectors mainly as scatterers and CdTe detectors as absorbers. The Compton camera contains three parts of detector clusters, namely Si detector part, CdTe bottom detector part, and CdTe side detector part (see Fig. \ref{fig:sgd-geometry}). Si detector part has 32 layers of Si detectors, each of which is pixelated into $16 \times 16$ pixels to enable it to detect the interaction position of photons. The thickness of a single Si layer is $600 {\rm\ \mu m}$, and accordingly the total thickness amounts to $1.92 {\rm\ cm}$, by which we obtain sufficiently high Compton scattering probability ($\sim 50 \%$ at $75 {\rm\ keV}$). CdTe bottom and side detector parts are respectively composed of 8 layers and $4 \times$ 2 layers of CdTe detectors. The CdTe detectors are arranged to surround the Si detector part. A CdTe bottom layer is pixelated into $16 \times 16$ pixels, whereas a CdTe side layer is pixelated into $16 \times 24$ pixels. The thickness of a CdTe layer is $750 {\rm\ \mu m}$. The size of each pixel is $3.2 {\rm\ mm} \times 3.2 {\rm\ mm}$ both for the Si and the CdTe layers. Their design details are shown in Watanabe et al. \cite{watanabe} and Tajima et al.\cite{tajima} \par
The charge signals generated in the detector material are amplified and converted to digital signals by 64-channel application-specific integrated circuits (ASICs). One ASIC has $8 \times 8$ channels, and hence a Si, CdTe bottom, and CdTe side layers are connected to 4, 4, and 6 ASICs, respectively. The total numbers of ASICs and channels are 208 and 13312, respectively.

\section{Experiment}
\label{sec:exp}

\begin{figure*}[thbp]
	\centering
	\includegraphics[width=0.9\hsize,keepaspectratio]{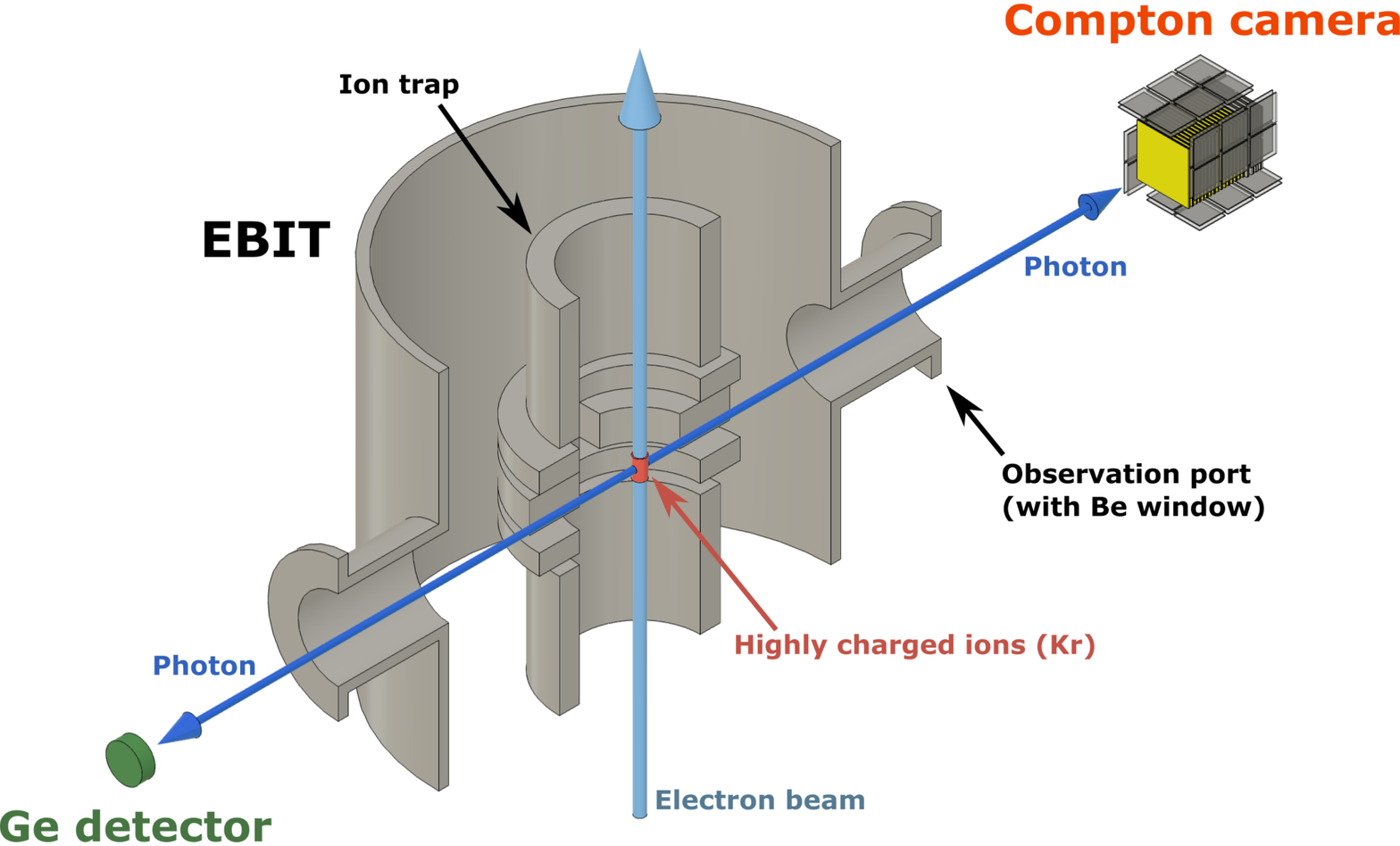}
	\caption{Schematic view of out experiment setup. The ion trap in the EBIT captures highly charged Kr ions, which are ionized by the electron beam. Radiative recombination x-rays emitted from the ions are transmitted to the Compton camera or to the Ge detector through the observation ports with Be windows. We employ the Ge detector and the EBIT-CC as a spectrometer and a polarimeter, respectively.}
	\label{fig:scheme-setup}
\end{figure*}
We performed an experiment to measure the degree of linear polarization of the $75{\rm\ keV}$ x-rays that are emitted through the radiative recombination to $n=1$ state from fully-ionized and hydrogen-like Kr ions. We explain the detail in this section. \par
We constructed the experiment apparatus by combining the EBIT-CC and a germanium (Ge) detector with the Tokyo Electron Beam Ion Trap (Tokyo-EBIT) \cite{tokyo-ebit, watanabe-ebit} at the University of Electro-Communications (UEC). A schematic view of the setup is shown in Fig. \ref{fig:scheme-setup}. Here, the EBIT-CC and the Ge detector are for polarimetry and for spectroscopy, respectively. \par
The Tokyo-EBIT generates fully-ionized and hydrogen-like Kr ions. It is composed of an electron gun and an ion drift tube. The electron beam travels along the $y$ axis of the Compton camera (see Fig. \ref{fig:sgd-geometry}) toward the $-y$ direction and ionizes Kr with electron collision. Highly charged ions generated by the beam are confined in the ion trap. The electron beam energy and current were $58.0 {\rm\ keV}$ and $130 {\rm\ mA}$, respectively. The typical electron beam radius is $30 {\rm\ \mu m}$\cite{herrmann} and the height (length in the $y$ direction) of the radiating area was evaluated to be smaller than $10 {\rm\ mm}$. The strength of the magnetic field in the trap region was $4 {\rm\ T}$. Kr gas was injected into the Tokyo-EBIT through a gas injector. The Kr ions were stored with an axial trapping potential of $100 {\rm\ V}$ and dumped every $30 {\rm\ s}$ to avoid the accumulation of contaminant heavy ions. The x-rays generated in the EBIT were radiated to detectors through Be windows, whose effective diameters are $50 {\rm\ mm}$ for the EBIT-CC and $25 {\rm\ mm}$ for the Ge detector. \par
The Compton camera was placed approximately $500 {\rm\ mm}$ away from the center of the EBIT. The operational temperature of the Compton camera was kept low at $-20{\rm\ {}^\circ C}$ to supress the leakage current of the detectors. The bias voltages were set to $230{\rm\ V}$ and $1000{\rm\ V}$ for the Si and CdTe detectors, respectively. To reduce noise, the energy threshold of the Si detectors was set to approximately $8 {\rm\ keV}$, which corresponds to the scattering angle $\theta_K \gtrsim 80^\circ$ where the total energy is $75 {\rm keV}$ (see Section \ref{sec:ana} for the definition of $\theta_K$). Then, most of the Compton scattered photons should be absorbed in the Si or CdTe-side part. The trigger for the CdTe-bottom part was disabled. The calibration procedure and the energy resolution of the Compton camera are described and discussed in Section \ref{sec:ana}. \par
The Ge detector was ORTEC GLP-36360/13 and was placed at the opposite side of the Compton camera with respect to the EBIT. The energy resolution of the Ge detector was $595 {\rm\ eV}$ ($0.73 {\rm\ \%}$) in full width at half maximum (FWHM) at $81 {\rm\ keV}$. The diameter and depth of its sensitive volume are $36 {\rm\ mm}$ and $13{\rm\ mm}$, respectively.

\section{Calibration, simulation, and data analysis}
\label{sec:ana}

In this section, we outline the method for data analysis. A raw event obtained with the EBIT-CC is composed of an array of ADC charge values $q_i$ and of pixel IDs ${\rm ID}_i$. From each set of hits $\{ (q_i, {\rm ID}_i) \, ; \, i = 1, 2, ..., N_{\rm hit} \}$ the event data is processed in the following steps. \par
First, $q_i$ are converted to the corresponding energy values $E_i$ by using calibration functions and ${\rm ID}_i$ are converted to their positions $\vec{x}_i$. We obtain the set $\{ (E_i, \vec{x}_i) \, ; \, i = 1, 2, ..., N_{\rm hit} \}$. \par
The second step is the event reconstruction, that is, mainly to determine the most probable sequence of the hits. We obtain the kinematic $\theta_K$, geometrical $\theta_G$ and the azimuthal scattering angle $\phi$, for each set. Here, $\theta_K$ is defined from the kinematics of Compton scattering as:
\begin{eqnarray}
	\cos \theta_K = 1 - m_e c^2 \left( \frac{1}{E - E_S} - \frac{1}{E} \right) \label{eq:theta-k}
\end{eqnarray}
where $E$ and $E_S$ denote the total photon energy and its energy deposit on the scatterer, respectively. The other parameter $\theta_G$ is defined as the angle between the momenta of the incoming photon and of the scattered photon (see Fig. \ref{fig:sgd-geometry}). \par
The third step is event selection. We impose several conditions on total energy, $\theta_K$, $\theta_G$, and hit sequence. Details are described in Subsection IV C. 

\subsection{Event reconstruction}
The event reconstruction, that is, to determine the ordering of hits $\{ (E_i, \vec{x}_i) \, ; \, i = 1, 2, ..., N_{\rm hit} \}$, is necessary to calculate $\theta_K$, $\theta_G$, and $\phi$. In this subsection we formulate a simple method of the reconstruction. \par
For the events of the total energy $E$ in a range of $73.0 {\rm\ keV} < E < 77.5 {\rm\ keV}$, $94.0{\rm\%}$ of the registered events containing Compton scattering are found to be two-hit events. Thus, we consider here only the two-hit Compton-scattered events and describe each of these events as $\{(E_1, \vec{x_1}), (E_2, \vec{x_2})\}$. With an aid of the Compton kinematics and the relation $2E \simeq 2 \times 75 {\rm\ keV} < m_e c^2 = 511 {\rm\ keV}$, the energy deposit on the scatterer $E_S$ is straightforwardly deduced to be smaller than that on the absorber $E_A$, as given by,  
\begin{eqnarray}
    E_A - E_S &=& 2 E_A - E \nonumber \\
    &=& E \left( \frac{2}{1 + \frac{E}{m_e c^2} (1 - \cos \theta)} - 1 \right) \nonumber \\
    &\geq& E \left( \frac{2}{1 + 2E / m_e c^2} - 1 \right) > 0 .
\end{eqnarray}
Therefore $E_S = \min \{ E_1, E_2 \}$ and $E_A = \max \{ E_1, E_2 \}$. 
The sequence of the two hits is also determined with this set of equations. \par
Once the event has been reconstructed, $\theta_K$ is calculated as in Eq. \ref{eq:theta-k}. In addition, $\theta_G$ and $\phi$ are computed as in Fig. \ref{fig:sgd-geometry}. The position of the trapped ions, $\vec{x}_{\rm source}$, which is necessary to compute $\theta_G$, is obtained with a Compton imaging technique\cite{takeda-mouse}. Here, we define two types of sequence: (1) {\it Si-Si sequence} where a photon is scattered in a Si detector and absorbed in another Si detector, and (2) {\it Si-CdTe-side sequence} where a photon is scattered in a Si detector and absorbed in a CdTe-side detector.

\subsection{Calibration}
To perform the event selection appropriately, precise energy calibration of the Compton camera is of particular importance. In this experiment, we optimized the settings of the ASICs for x-rays with the energy $\sim 75 {\rm\ keV}$, which is lower than those adopted for the {\it Hitomi} SGD. \par
The procedure of calibrating the Compton camera consists of three steps. First, the non-uniformity of the outputs from the ADCs for each channel within an ASIC was evaluated by inputting constant charge pulses called {\it test pulse} into each channel. Then the outputs from the ADCs within an ASIC could be treated in the same manner regardless of the channel. Second, the conversion factors from the ADC value to energy were calculated using point sources of radio isotopes. Third, the energy outputs lower than the trigger threshold were corrected for each ASIC connected to a Si detector. \par
We performed the calibration, using gamma-rays from a ${}^{133}{\rm Ba}$ source, and found the energy resolution of the Compton camera for reconstructed Compton events to be $4.7 {\rm\ keV}$ (5.8\,\%) in FWHM at $E_S + E_A \simeq 81 {\rm\ keV}$ after the event selection (see Subsection IV C). Note that this value was calculated by combining the outputs from all the enabled channels, not from a single channel.

\subsection{Event selection}
\begin{figure}
    \centering
    \includegraphics[width = \linewidth, keepaspectratio]{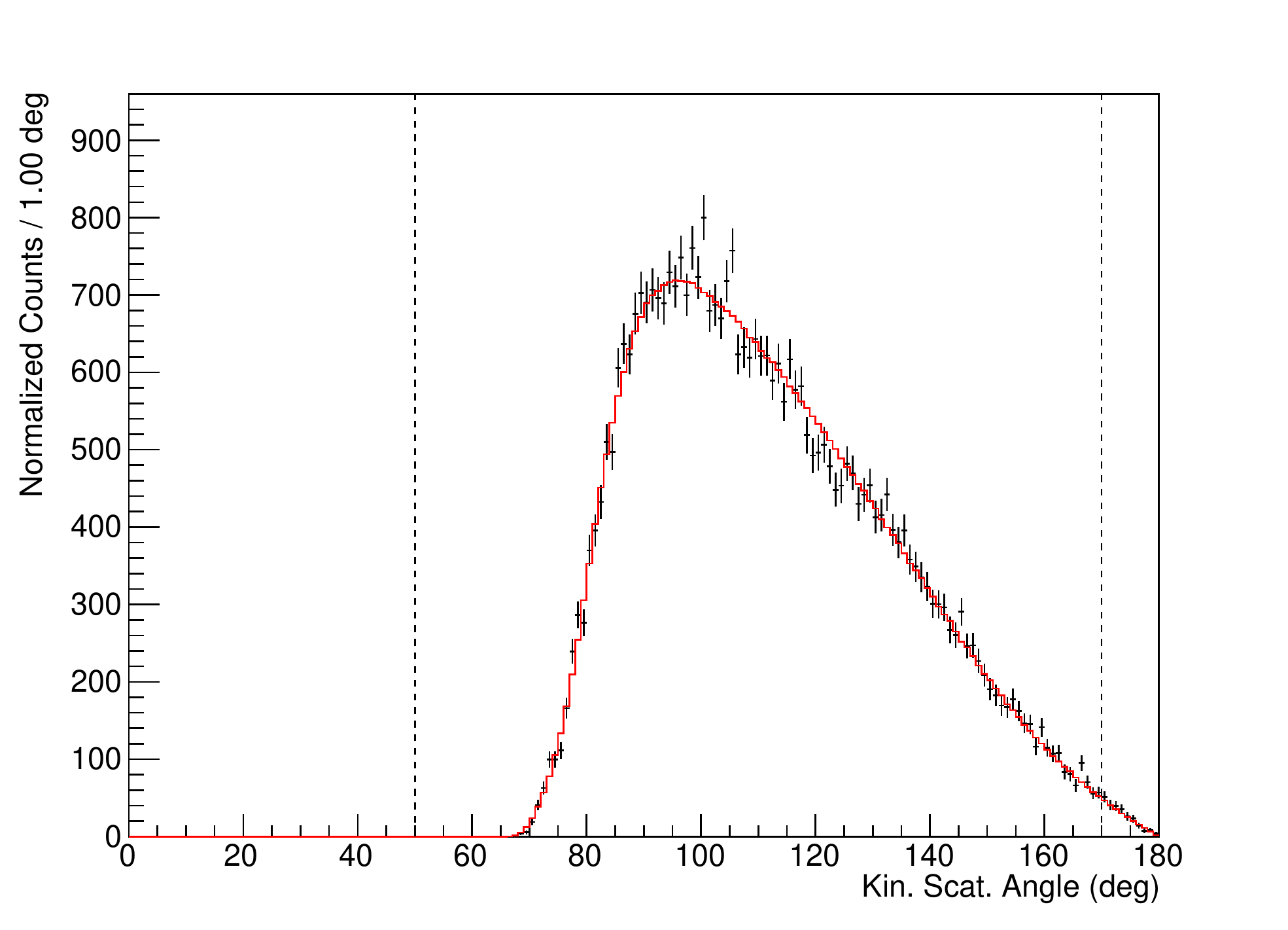}
    \caption{Distributions of $\theta_K$ of the events that satisfy all the selection criteria except for (3) (see text). Data points (black) show the Kr measurement minus the background. Solid line (red) shows the simulation. Dotted lines show criterion (3).}
    \label{fig:hist-thk}
\end{figure}
\begin{figure}
    \centering
    \includegraphics[width = \linewidth, keepaspectratio]{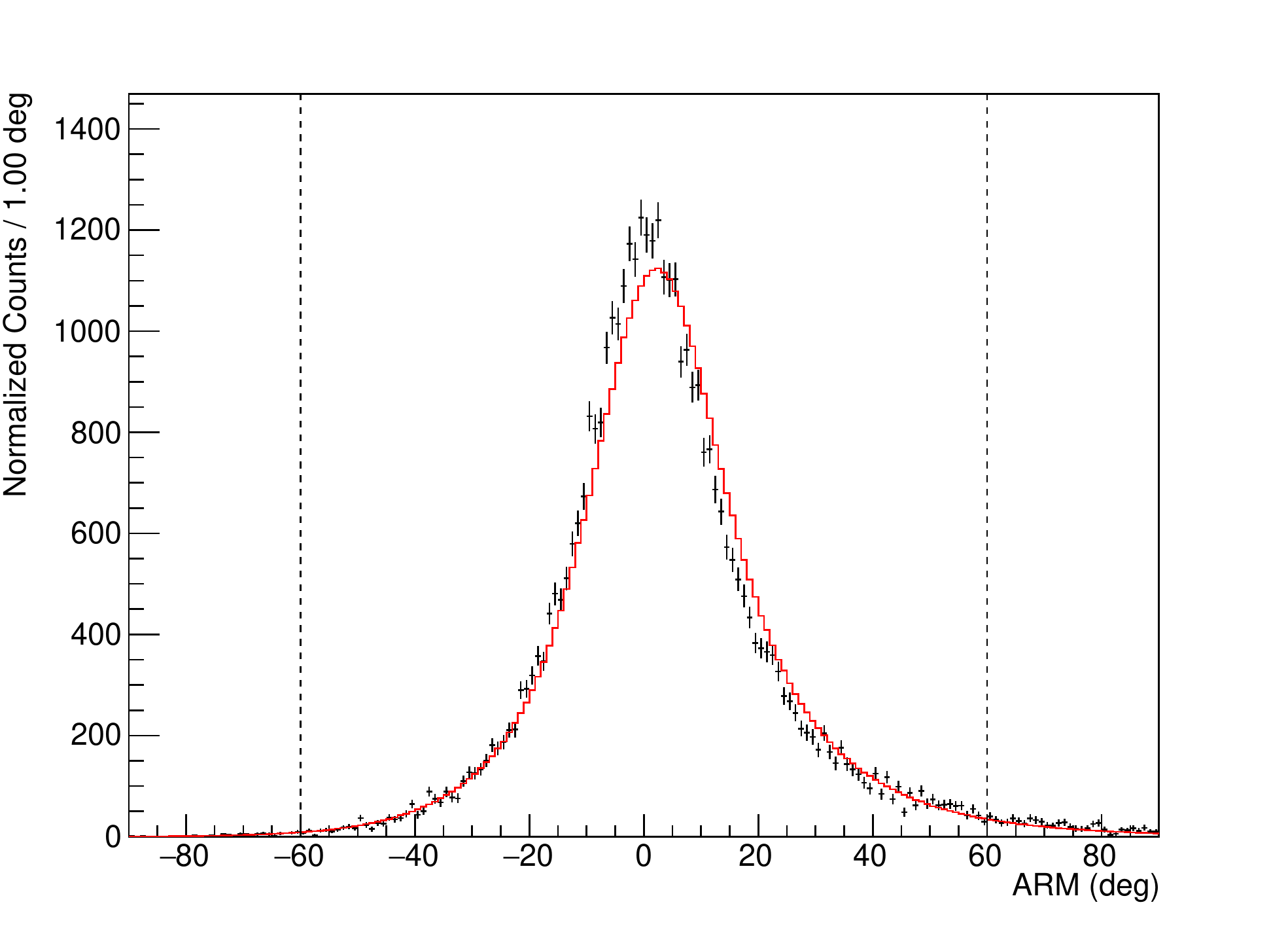}
    \caption{ARM distributions of the events that satisfy all the selection criteria except for (4) (see text). Data points (black) show the Kr measurement minus the background. Solid line (red) shows the simulation. Dotted lines show criterion (4). Discrepancies between the data and the simulation could result from the energy-spectral distortion (low-energy tail) of the CdTe detectors which could not be simulated.}
    \label{fig:hist-arm}
\end{figure}
\begin{figure*}
    \centering
    \subfigure[]{
        \includegraphics[width = 0.45\linewidth, keepaspectratio, page = 1]{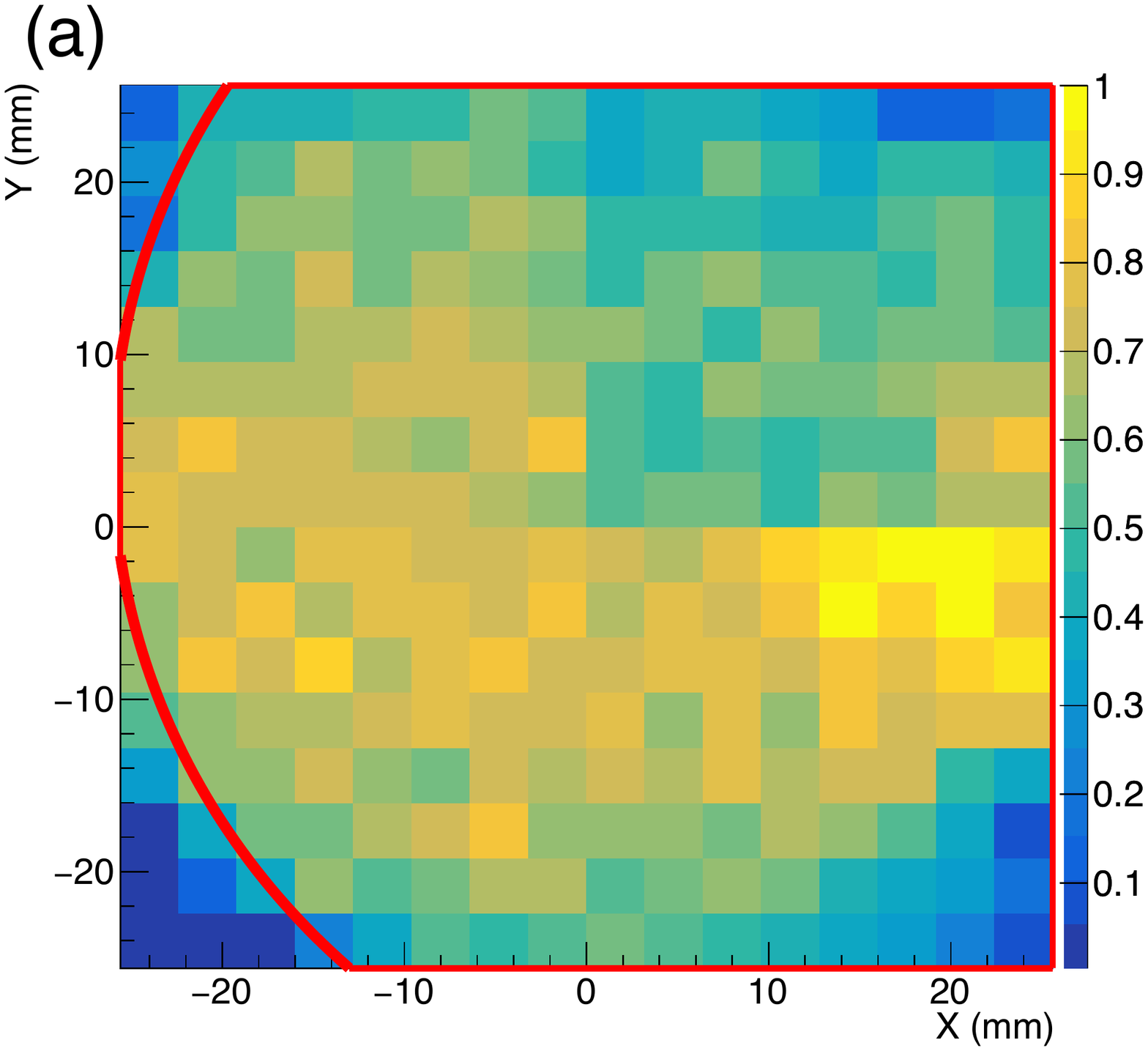}
    }
    \subfigure[]{
        \includegraphics[width = 0.45\linewidth, keepaspectratio, page = 2]{./hist_pos2d_20210305.pdf}
    }
    \caption{Distributions of first-hit positions of the events that satisfy all the selection criteria except for (5) (see text). (a) and (b) show the Kr measurement minus the background and the simulation, respectively. Both distributions are normalized such that the maximum bin content is equal to $1$. Some pixels at the bottom-left in (a) contain less events than those in (b). The top-right quarter parts in the both distributions contain less events than the other parts because we disabled a significant number of Si layers located there for their electric device troubles. Red lines show criterion (5).}
    \label{fig:hist-pos2d}
\end{figure*}
We determine the criteria for the event selection by evaluating the hit sequence, total energy, $\theta_G$, and angular resolution measure ${\rm (ARM)} = (\theta_K - \theta_G)$, and also considering the distribution of the first hit positions of the events as follows.
\begin{enumerate}
	\item[(1)] It is either a Si-Si sequence or Si-CdTe-side sequence.
	\item[(2)] Total energy $E$ satisfies the following conditions.
	\begin{enumerate}
	    \item[(2-a)] $79.0 {\rm\ keV} < E < 83.0 {\rm\ keV}$ for the ${\rm Ba}$ data.
	    \item[(2-b)] $73.0 {\rm\ keV} < E < 77.5 {\rm\ keV}$ for the ${\rm Kr}$ data.
	\end{enumerate}
	These threshold values correspond to the FWHM of the peak in question.
	\item[(3)] Geometrical scattering angle $\theta_K$ satisfies $50.0^\circ < \theta_K < 170.0^\circ$. Fig. \ref{fig:hist-thk} shows this condition.
	\item[(4)] $-60.0^\circ < {\rm ARM} < 60.0^\circ$ (see Fig. \ref{fig:hist-arm}).
	\item[(5)] First hit position $(x_1, y_1)$ satisfies $(x_1 - 14.25{\rm\ mm})^2 + (y_1 - 4.0{\rm\ mm})^2 < (40.25{\rm\ mm})^2$. We find that the number of first-hit events is significantly smaller in some regions in distribution plots of first-hit positions (bottom-left pixels in Fig. \ref{fig:hist-pos2d}). This could be caused by some assemblies in front of the EBIT-CC. We impose this condition to exclude such regions.
\end{enumerate}
If an event does not satisfy all the conditions above, the event is excluded.

\subsection{Monte Carlo simulation}
\begin{figure}
	\centering
	\includegraphics[width=1.0\linewidth,keepaspectratio]{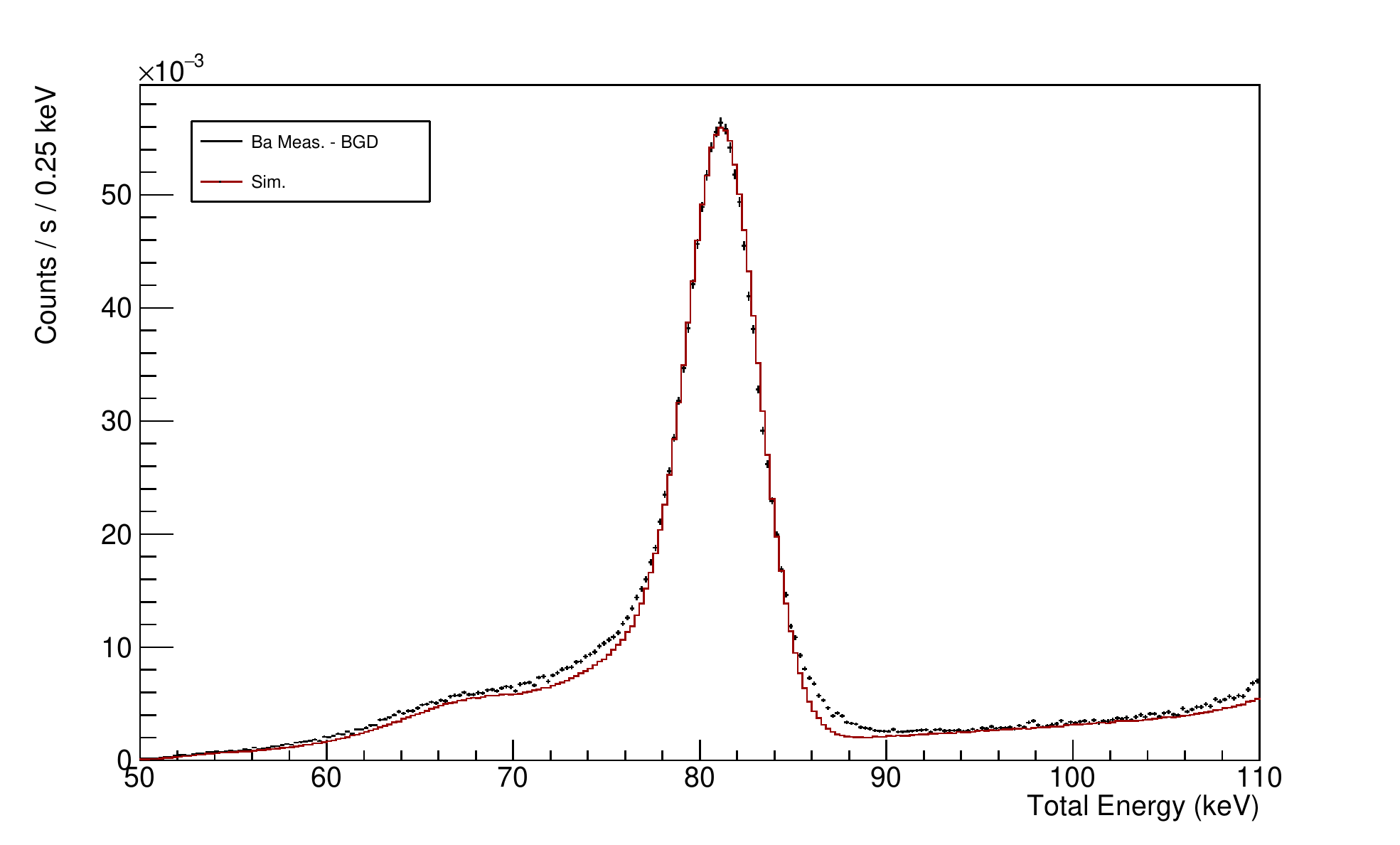}
	\caption{Measured background-subtracted energy spectra (in black) of the gamma rays from a ${}^{133}{\rm Ba}$ point source, overlaid with the simulation (red solid line). The used events are of the Si-Si and Si-CdTe-side sequences.}
	\label{fig:hist-energy-ba}
\end{figure}
\begin{figure}
	\centering
	\includegraphics[width=1.0\linewidth,keepaspectratio]{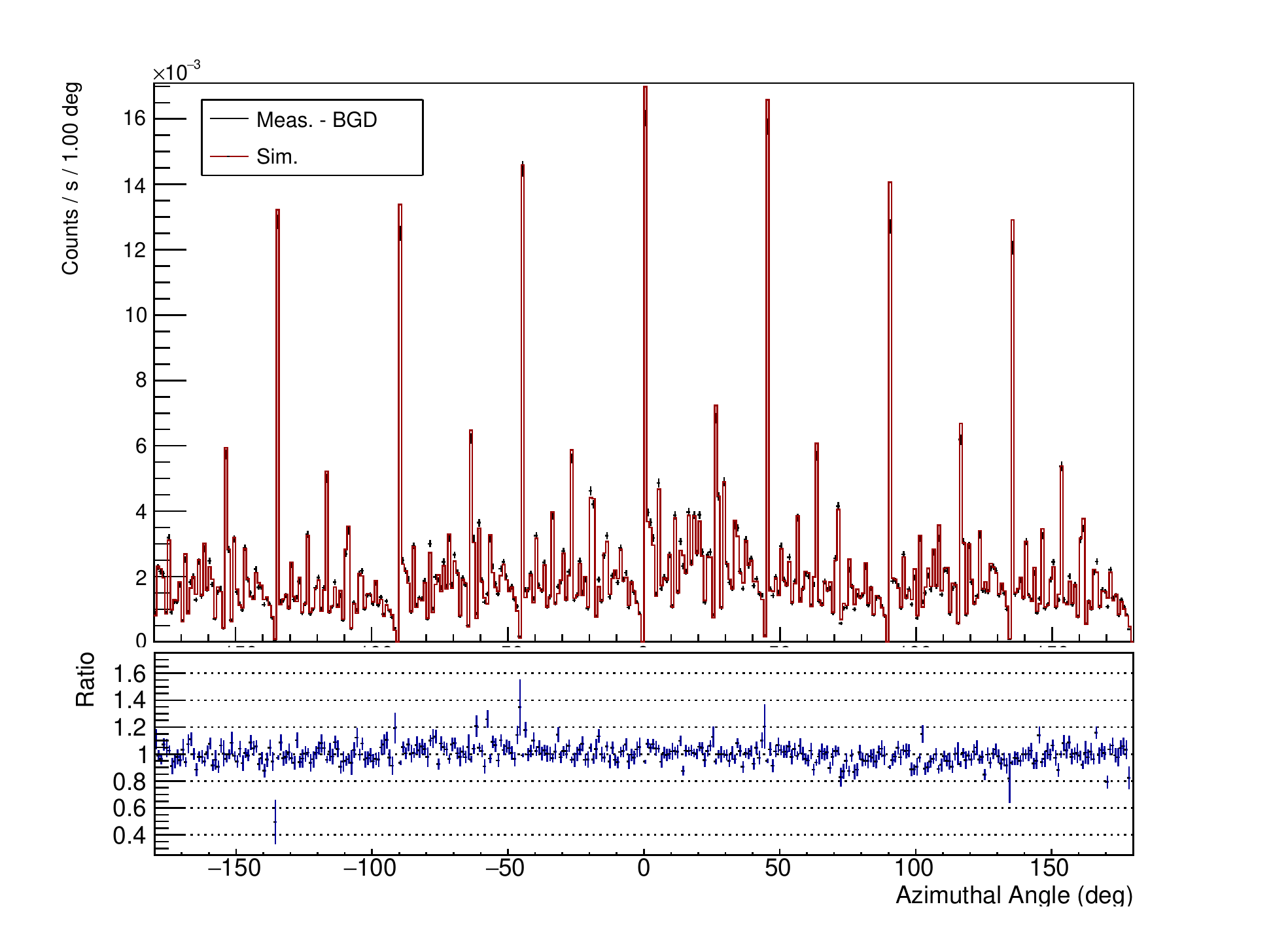}
	\caption{Measured azimuthal angular distribution (in black) for the gamma rays from a ${}^{133}{\rm Ba}$ point source, overlaid with the simulation (red solid line). The used events are of the Si-Si and Si-CdTe-side sequences and their energy range is $79.0 {\rm\ keV} - 83.0 {\rm\ keV}$. The background is subtracted. Lower panel shows the ratio (in blue) of the measurement to simulation.}
	\label{fig:hist-ph-ba}
\end{figure}
\begin{figure}
    \centering
    \includegraphics[width=1.0\linewidth,keepaspectratio]{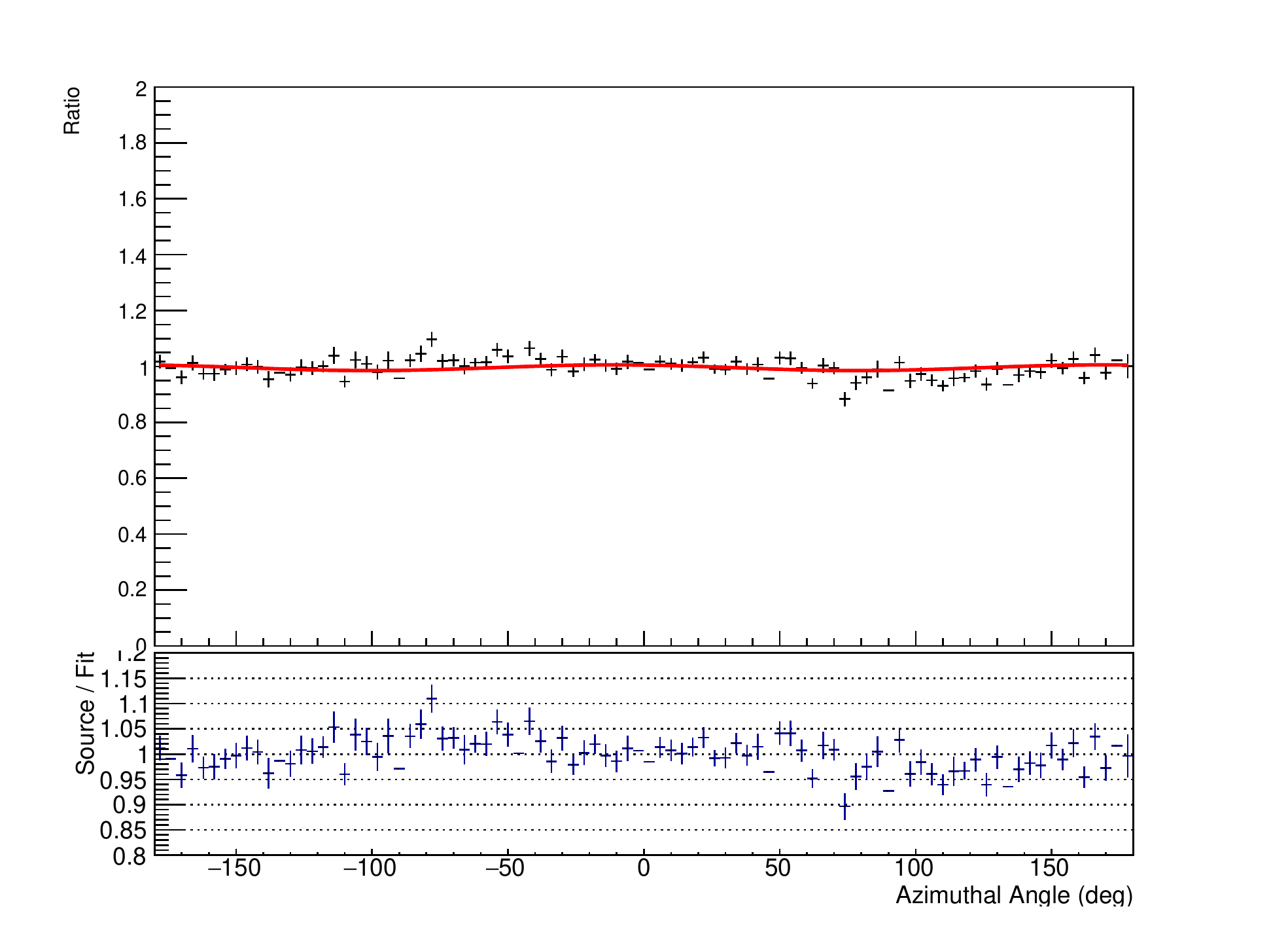}
    \caption{Modulation curve for the gamma rays from a ${}^{133}{\rm Ba}$ source, overlaid with the best-fit model function (red solid line), which is a linear superposition of a constant and $\cos (2\phi - {\rm const.})$. The background is subtracted from the data. Lower panel shows the ratio (in blue) of the data to best-fit result.}
    \label{fig:curve-ba}
\end{figure}
We performed a series of Monte Carlo simulations to obtain the azimuthal distribution model $d_i^{\rm (model)}$ (see Sec. \ref{sec:cc} for the context and notations). We utilized the {\ttfamily ComptonSoft} simulation toolkit\cite{comptonsoft} for the simulations. {\ttfamily ComptonSoft} can simulate the physical processes in the detector including Compton scattering, the photoelectric effect, and the Doppler broadening\cite{ordonez,zoglauer}. It employs the {\ttfamily Geant4} simulation framework\cite{geant4-agostinelli,geant4-allison} for physical calculations. \par
Since the EBIT-CC is a multi-layer, pixelated detector, the detector response has some non-uniformity with respect to the position of scattering $\vec{x_S}$. This non-uniformity could result from anisotropy of the radiation from sources, since we assume isotropic radiation in the simulation. To compensate for the non-uniformity, we split simulated events into $16$ divisions with respect to the scattering position projected onto the $xy$ plane $(x_S, y_S)$. Since they satisfy $-25.6 {\rm\ mm} < x_S < +25.6 {\rm\ mm}$ and $-25.6 {\rm\ mm} < y_S < +25.6 {\rm\ mm}$, we made $4$ divisions with the area of $12.8 {\rm\ mm} \times 12.8 {\rm\ mm}$ along both the $x$ and $y$ directions and multiply a scaling factor $\nu_j\ (j = 1, 2, ..., 16)$ for each division. The content of the $i$-th bin of the energy spectrum $e_{i'}$ is calculated as
\begin{eqnarray}
    e_{i'} = \sum_{j=1,2,...,16} \nu_j e_{i'j} ,
\end{eqnarray} 
where $e_{i'j}$ is the content of the $i$-th bin of $j$-th energy spectrum. The distribution of $\phi$ is calculated in the same manner. Specifically, $\nu_j$ are calculated by fitting the energy spectra. \par
To verify the simulation parameters, we simulated gamma-rays radiated from a ${}^{133}{\rm Ba}$ point source and compared the simulation result with the measured one with the EBIT-CC. The measured and simulated energy spectra of the gamma-rays are shown in Fig. \ref{fig:hist-energy-ba}. We find the two spectra to agree well. Figs. \ref{fig:hist-ph-ba} and \ref{fig:curve-ba} compare the azimuthal angular distributions and modulation curves, respectively, from the actual measurement and simulation. We find from the latter a small modulation proportional to $\cos (2 \phi - {\rm const.})$, which implies that the gamma-rays from the ${}^{133}{\rm Ba}$ source are unpolarized. In Sec. \ref{sec:res}, we discuss systematic uncertainties on the degree of polarization using this curve.

\section{Results and discussion}
\label{sec:res}

\subsection{Results}
The count rate of the all registered events was $\sim 2.3 \times 10^2 {\rm\ s^{-1}}$ both for the Ge detertor and the EBIT-CC. Among $3.8 \times 10^7$ events recorded with the EBIT-CC, $4.3 \times 10^4$ events ($0.11\%$) were selected.

\subsubsection{Energy spectra (comparison with those of the Ge detector)}
\begin{figure}[t]
	\centering
	\includegraphics[width=1.0\linewidth,keepaspectratio]{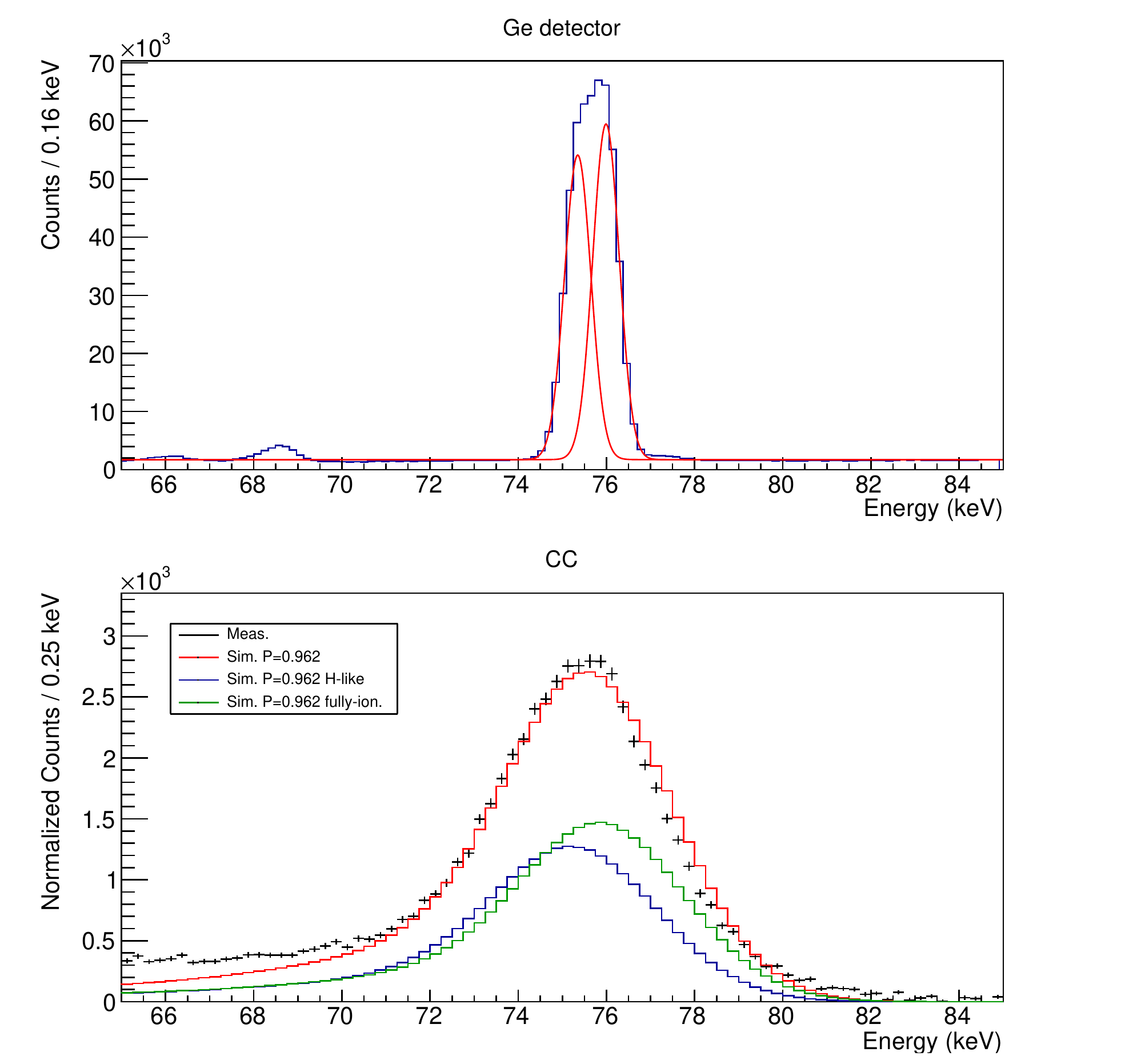}
	\caption{Energy spectra of the Kr measurements. (Top panel) Spectrum with the Ge detector. Red lines show the derived peaks (see text). (Bottom panel) The spectrum with the EBIT-CC for the events of the Si-Si and Si-CdTe-side sequences. Black dots show the measured data minus the background. Red line shows the simulation for $\Pi=0.962$ (the best fit). Blue and green lines show the simulated hydrogen-like and fully-ionized Kr lines, respectively.}
	\label{fig:hist-energy-kr}
\end{figure}
\begin{figure}
    \centering
    \subfigure[Kr measurement.]{
        \includegraphics[width=0.9\linewidth,keepaspectratio,page=1]{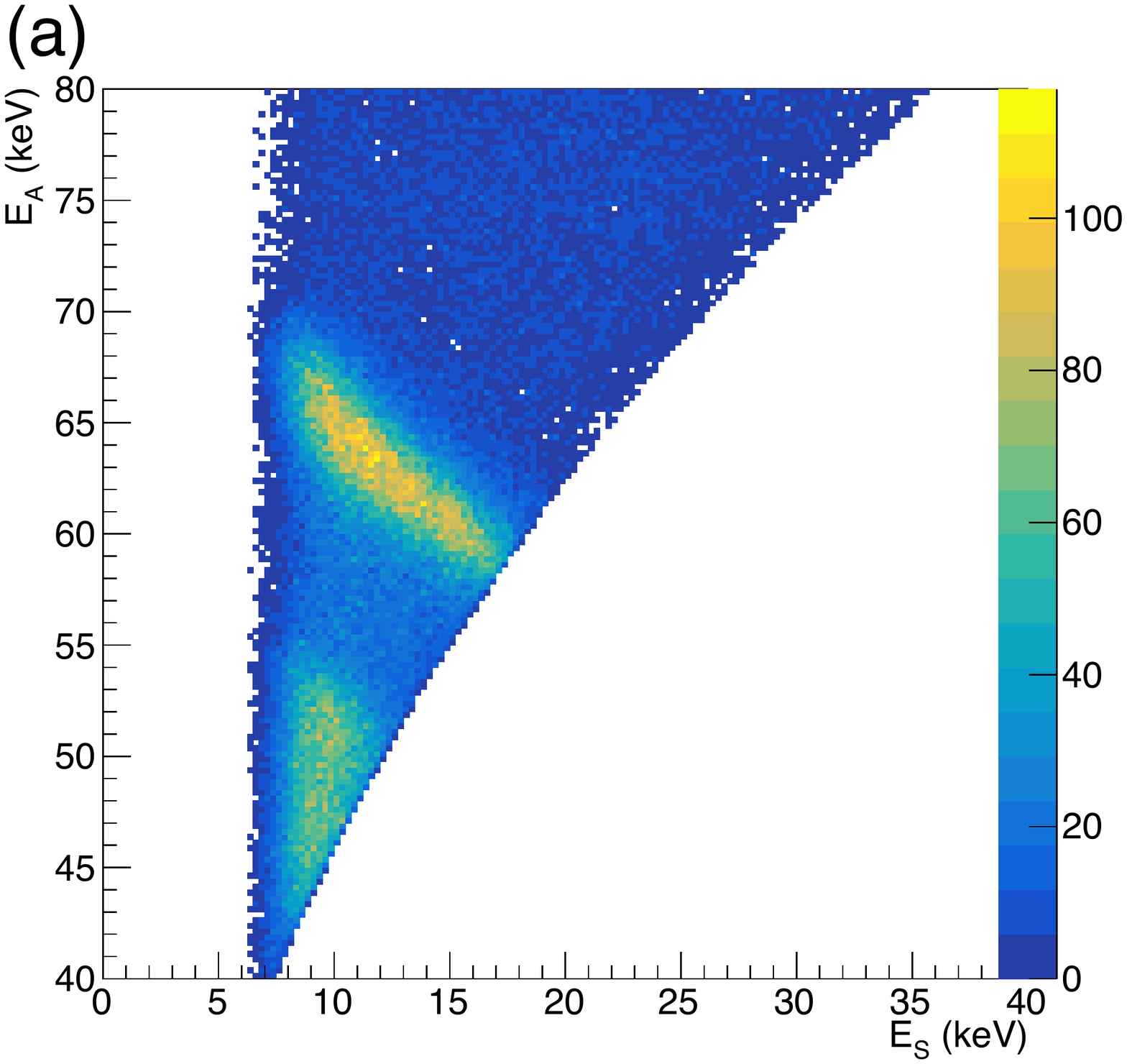}
    }\\
    \subfigure[Background measurement.]{
        \includegraphics[width=0.9\linewidth,keepaspectratio,page=2]{./hist_e1e2_16divk_kr_total_20201227_ml.pdf}
    }
    \caption{Two-dimensional histograms of the energy deposit on the scatterer $E_S$ versus energy deposit on the absorber $E_A$. (a) is for the Kr measurement and (b) for the background measurement. The histograms for the Kr and background measurements are normalized with respect to the Kr measurement.}
    \label{fig:hist-e1e2-kr}
\end{figure}
The energy spectra obtained with the Ge detector and Compton camera are shown in Fig. \ref{fig:hist-energy-kr}. For the latter, the events of the Si-Si and Si-CdTe-side sequences are used. \par
Using Fig. \ref{fig:hist-energy-kr}, we evaluate the calibration error. The radiative recombination lines of Kr ions is clearly present at around $75 {\rm\ keV}$ in both spectra. This peak contains two components, namely radiative recombination lines into fully-ionized and hydrogen-like Kr. The difference in energy of the two is $0.640 {\rm\ keV}$\cite{nakamura-rhodium}. By fitting the peak of the spectrum from the Ge detector with two Gaussian functions whose peak energies differ by $0.640 {\rm\ keV}$, the peak energies are derived to be $75.35 \pm 0.03 {\rm\ keV}$ (hydrogen-like) and $75.99 \pm 0.03 {\rm\ keV}$ (fully-ionized), which as a whole correspond to the peak at $75.5\pm0.2 {\rm\ keV}$ for the Compton camera. The intensity-weighted average of the two peak positions in the Ge-detector spectrum is $75.68 \pm 0.02 {\rm\ keV}$, which is consistent with the peak position in the Compton-camera spectrum. \par
The two-dimensional histograms of the energy deposit on the scatterer $E_S$ versus energy deposit on the absorber $E_A$ are plotted in Fig. \ref{fig:hist-e1e2-kr}. The lower bound at around $E_S \sim 7 {\rm\ keV}$ of each two-dimensional histogram results from the energy threshold of the Si detectors. The Kr-measurement histogram shows a stripe pattern that roughly follows $E_S + E_A \simeq 75 {\rm\ keV}$, which should correspond to the spectral peak in Fig. \ref{fig:hist-energy-kr}.

\subsubsection{$\phi$ distributions and modulation curves}
\begin{figure}[htbp]
    \centering
    \includegraphics[width=0.9\linewidth,keepaspectratio]{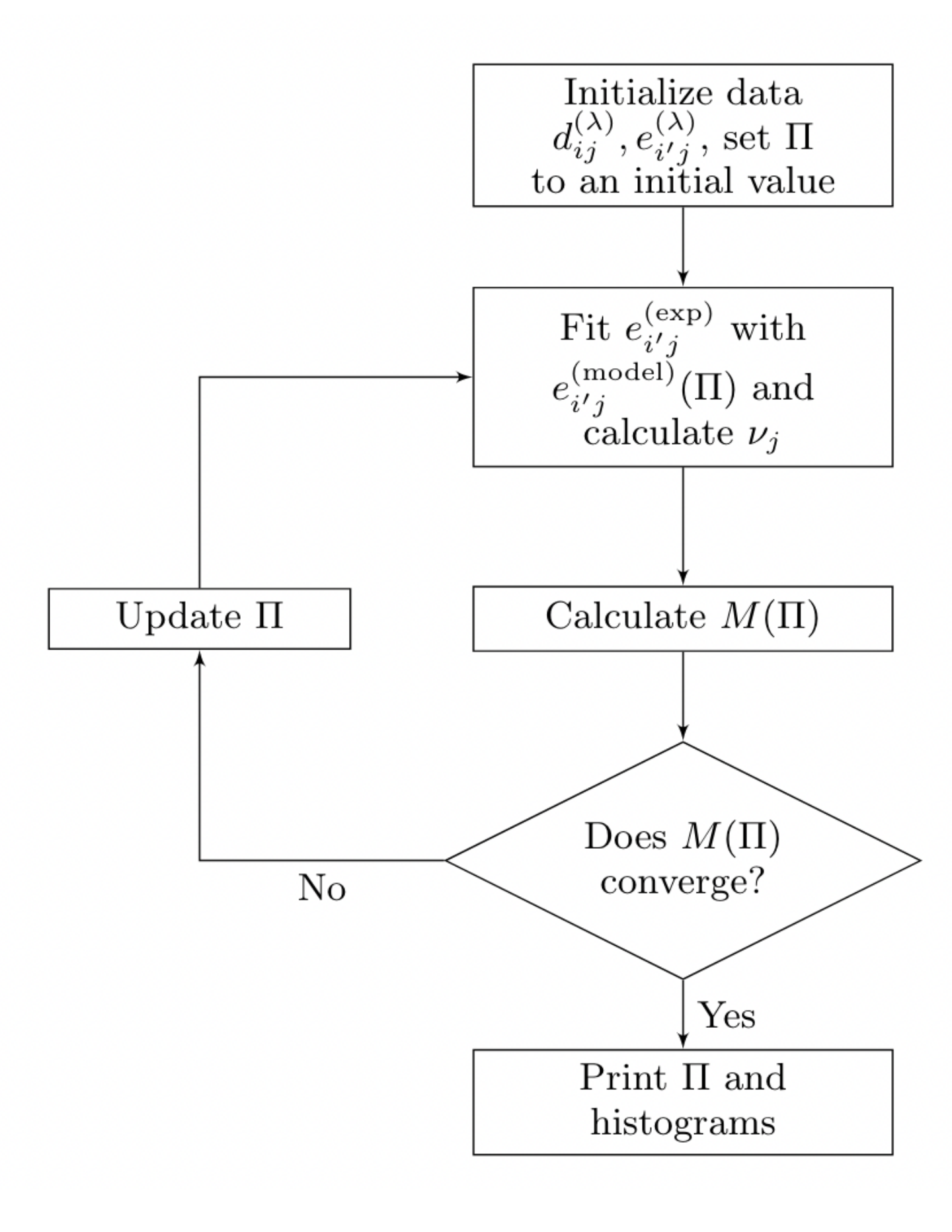}
    \caption{Schematic diagram of the procedure to find the optimal degree of polarization $\Pi$. Note $\lambda = {\rm exp}, {\rm bg}, {\rm 0}, {\rm or\ } {\rm 100}$. Definitions of $d_{ij}$, $e_{i'j}$, $\nu_j$, and $M( \Pi )$ are presented in the text.}
    \label{fig:diagram-fit}
\end{figure}
\begin{figure}[htbp]
    \centering
    \includegraphics[width=1.0\linewidth,keepaspectratio]{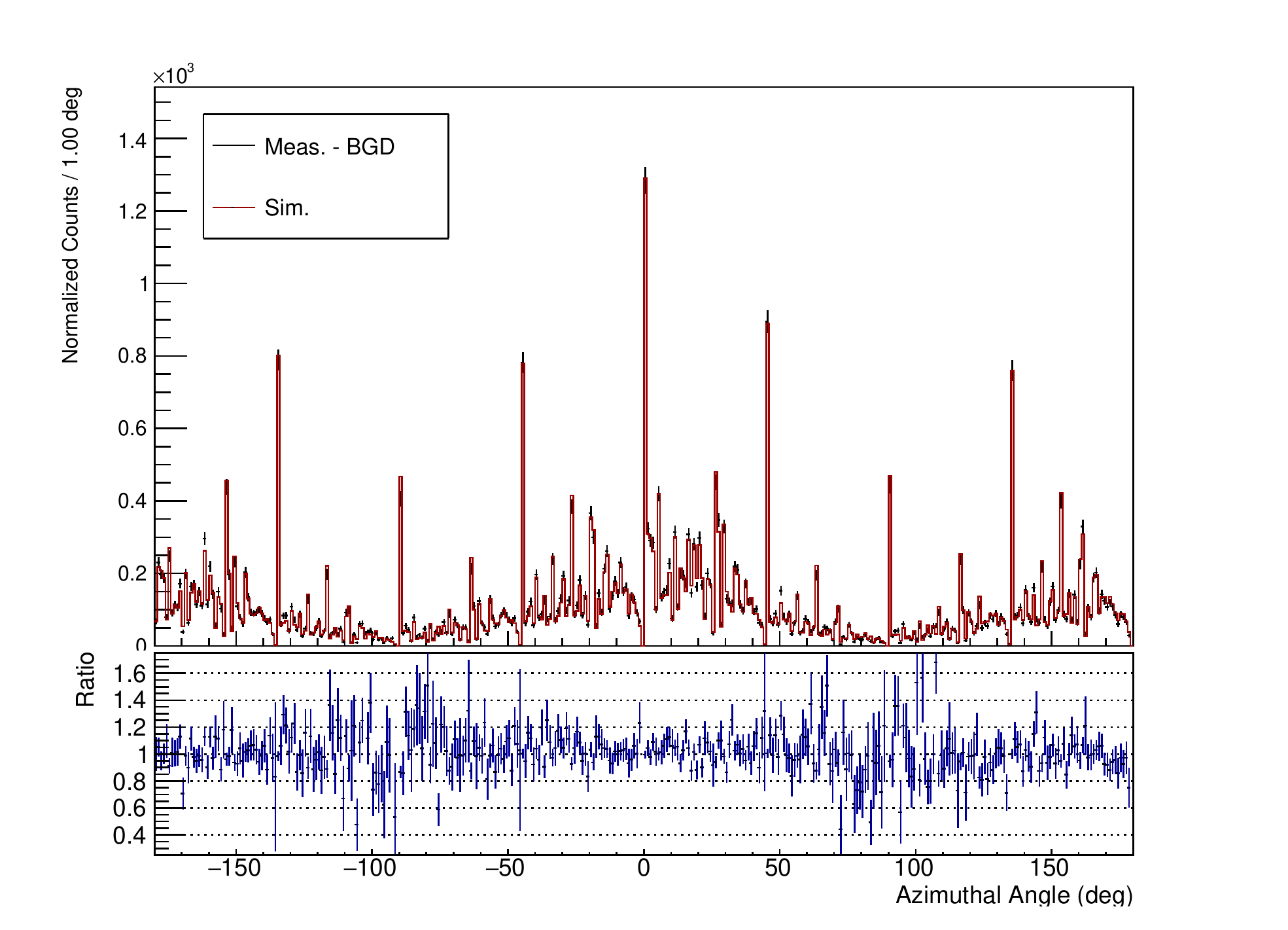}
    \caption{Azimuthal angular distribution of the ${\rm Kr}$ measurement (in black), overlaid with the simulation result for $\Pi=0.962$ (the best-fit result) in the red histogram. The events of the Si-Si and Si-CdTe-side sequences are selected. Lower panel shows the ratio (in blue) of the measured data to simulation.}
    \label{fig:hist-ph-kr}
\end{figure}
\begin{figure}[htbp]
    \centering
    \includegraphics[width=1.0\linewidth,keepaspectratio]{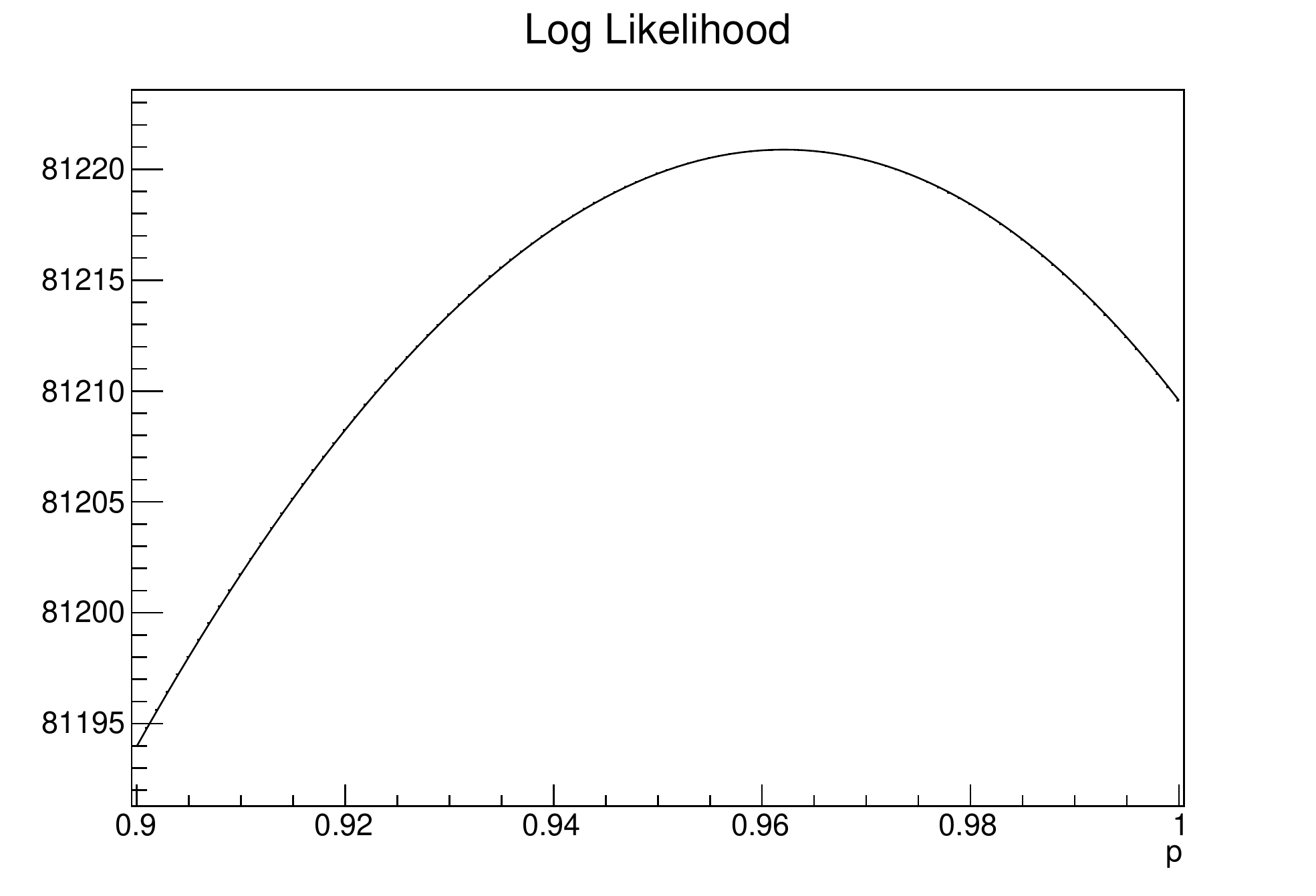}
    \caption{Logarithmic likelihood $M( \Pi )$. The maximum value is found at $\Pi = 0.962$.}
    \label{fig:gr-ll}
\end{figure}
\begin{figure}[htbp]
    \centering
    \includegraphics[width=1.0\linewidth,keepaspectratio]{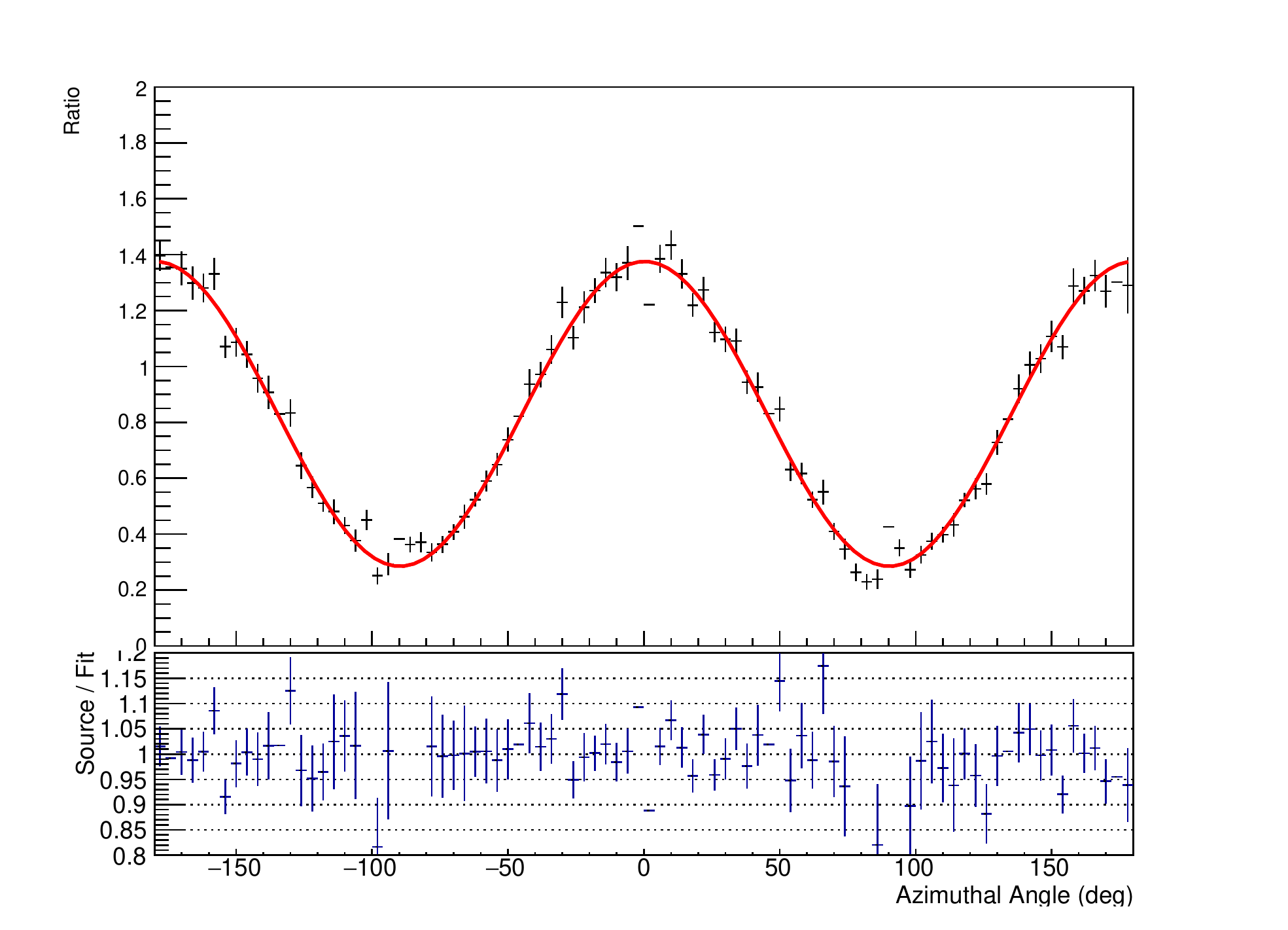}
    \caption{Modulation curve for the ${\rm Kr}$ radiative recombination, where the background is subtracted. Red line shows the best-fit model function, which consists of a constant and $\cos (2\phi - {\rm const.})$. Lower panel shows the ratio (in blue) of the data to best-fit result.}
    \label{fig:curve-kr}
\end{figure}
The degree of polarization, $\Pi$, of the detected x-rays is derived as follows. First, we construct $\phi$ histograms (designated as $d_{ij}$) and energy histograms ($e_{i'j}$) from the data of the ${\rm Kr}$ measurement, background measurement, and a pair of simulations on the basis of $100{\rm\%}$ polarized and unpolarized x-rays. Here, $i'$ and $i$ denote the bin number and $j$ the index of division (see Section \ref{sec:ana}). The background data ($e_{i'j}^{({\rm bg})}$ and $d_{ij}^{({\rm bg})}$) are normalized such that the observation time is equal to that of the ${\rm Kr}$ data. To obtain the normalization factors $\nu_j$ for the simulation data, we fit the energy histograms of the ${\rm Kr}$ experiment ($e_{ij}^{({\rm exp})}$) with the model ($e_{i'j}^{({\rm model})}$) for each of $i' = 1, 2, ..., {N'}_{\rm bin}$ and $j = 1, 2, ..., 16$. The model is given by
\begin{equation}
    e_{i'j}^{({\rm model})} ( \Pi ) = e_{i'j}^{({\rm bg})} + \nu_{j} \left( (1 - \Pi ) e_{i'j}^{({\rm 0})} + \Pi e_{i'j}^{({\rm 100})} \right) , \label{eq:e-model}
\end{equation}
where the superscripts $(0)$ and $(100)$ denote the simulation data for unpolarized and $100{\rm\%}$-linearly-polarized x-rays, respectively. We compute the logarithmic likelihood ($M( \Pi )$) for the comparison between $d_{ij}^{({\rm exp})}$ and $d_{ij}^{({\rm model})}$ with fixed $\nu_j$, as given by
\begin{equation}
    M ( \Pi ) = \sum_{ij} \left( d_{ij}^{({\rm exp})} \ln{d_{ij}^{({\rm model})}} - d_{ij}^{({\rm model})} \right) + {\rm const.} \label{eq:ll}
\end{equation}
where
\begin{equation}
    d_{ij}^{({\rm model})} ( \Pi ) = d_{ij}^{({\rm bg})} + \nu_{j} \left( (1 - \Pi ) d_{ij}^{({\rm 0})} + \Pi d_{ij}^{({\rm 100})} \right) \label{eq:d-model}
\end{equation}
for a range of $i = 1, 2, ..., N_{\rm bin}$. In Eq. (\ref{eq:ll}), we assume that $d_{ij}^{({\rm exp})}$ obey Poisson distributions with the mean $d_{ij}^{({\rm model})}$. We iterate the procedure updating the $\Pi$ value until $M( \Pi )$ reaches its maximum. Fig. \ref{fig:diagram-fit} gives the block diagram that summarizes the procedure. \par
Fig. \ref{fig:hist-ph-kr} shows the resultant summed histograms of $\phi$ with the best-fit result. We find that $M( \Pi )$ takes its maximum at $\Pi = 0.962$ ($0.953$-$0.970$; $1\sigma$ confidence). Fig. \ref{fig:gr-ll} shows the derived logarithmic likelihood curve. \par
Fig. \ref{fig:curve-kr} shows the obtained modulation curve. We find $\Pi = 0.968 \pm 0.010 \ (1\sigma )$ by fitting the curve with the function
\begin{eqnarray}
	\frac{d^{\rm (exp)} (\phi) - d^{\rm (bg)} (\phi)}{d^{\rm (0)} (\phi)} = A \left( 1 + Q \cos 2 \left( \phi - \phi_0 \right) \right)	, \label{eq:modulation}
\end{eqnarray}
where $A$, $Q$, and $\phi_0$ are the fitting parameters; $Q$ is the so-called modulation factor.
Since the modulation factor is proportional to the degree of polarization, we obtain $\Pi = Q / Q_{100}$ where $Q_{100}$ is the modulation factor for $d^{\rm (100)} / d^{\rm (0)}$.

\subsection{Discussion}
We evaluate the systematic uncertainty in the Monte Carlo simulation. In Fig. \ref{fig:curve-ba}, we find the modulation factor $Q = 0.010$, which would correspond to $\Pi \sim 0.015$ under an assumption of $Q_{100} \sim 0.7$. Then, we interpret this value as the uncertainty of the simulation, $\Delta \Pi \sim 0.015$. \par
We conjecture that the systematic uncertainty resulting from the event selection is relatively small. We find that $\Pi$ is fairly insensitive to the threshold of $\theta_K$ and ARM. Specifically, the optimal value of $\Pi$ varies from $0.958$ to $0.962$ when we vary the threshold of $\theta_K$ for ranges of $0^\circ < \min\, \theta_K < 70^\circ$ and $130^\circ < \max\, \theta_K < 170^\circ$, and from $0.962$ to $0.968$ when we vary the threshold of ${\rm ARM}$ from $\pm 60^\circ$ to $\pm 10^\circ$. These variations are smaller than the statistical uncertainty of $\Pi$. The uncertainty resulting from the detector non-uniformity (see Sec. \ref{sec:ana}) is within the statistical errors. \par
We find the systematic uncertainty resulting from the experimental setup to be sufficiently small. The spiral motion of the electron beam could not result in significant uncertainties, because the angle between the beam axis and the momentum of electrons is evaluated to be smaller than $\sim 4^\circ$. The misalignment between the directions of the x-rays and the azimuth of the Compton camera could not be larger than $1^\circ$. The size of the x-ray emitting region, approximately $60 {\rm\ \mu m} \times 10 {\rm\ mm}$, would be small enough to be treated as a point source, in comparison with the angular resolution of the EBIT-CC. \par
Given the systematic uncertainties discussed above, we conclude $\Pi = 0.962 \pm 0.008\ {\rm (statistical)}\ \pm 0.015\ {\rm (systematic)}$, or $\Pi = 0.962 \pm 0.023$ if we combine the statistical and the systematic uncertainties. In other words, we have obtained the degree of polarization with an error of $\sim 2 {\rm \%}$.\par
The method that we have presented in this paper is applicable also to other line emissions, such as the dielectronic recombination x-rays of heavy ions. For example, it is theoretically expected that the polarization degree of dielectronic recombination x-rays depends on the zero-frequency approximation in the Breit interaction\cite{fritzsche, tong, nakamura-evidence}. To determine the limit of the approximation, one would need an absolute precision of $\Delta \Pi \sim 0.02$ for highly charged bismuth (Bi) ions, for the expected $\Pi \sim 0.2$. It would be reasonable to conclude that these criteria can be achieved by applying our method.


\section{Summary and conclusion}
\label{sec:sum}

We developed the EBIT-CC, a novel Compton polarimeter with Si/CdTe semiconductor detectors, and attached it on the Tokyo-EBIT in the UEC. An experiment was performed to evaluate its polarimetric capability through an observation of radiative recombination x-rays emitted from highly charged Kr ions that was generated by the Tokyo-EBIT. The Compton camera was calibrated for the $\sim 75 {\rm\ keV}$ x-rays. We developed event reconstruction and selection procedures and applied them to every registered event. As a result, we obtained the degree of polarization $\Pi = 0.962 \pm 0.023$. This precision would be satisfactory for evaluation of the QED and relativistic effect, for which $\Delta \Pi \sim 0.02$ for $\Pi \sim 0.2$ is required. To conclude, we have established a new polarimetric method using the EBIT-CC, which shows sufficient polarimetric capability and sensitivity for high-$Z$ atomic x-ray observations. 

\section*{Acknowledgements}

This work was supported by JSPS KAKENHI Grant-in-Aid for Scientific Research on Innovative Areas `Toward new frontiers: Encounter and synergy of state-of-the-art astronomical detectors and exotic quantum beams' 18H05458, 18H05463 and 19H05187, Grant-in-Aid for Scientific Research (A) 20H00153, World Premier International Research Center Initiative (WPI), MEXT, Japan, and JST-SENTAN program (Development of an advanced gamma-ray imaging system with an ultra-wide field of view and a high sensitivity).
\section*{Data Availability}
The data that support the findings of this study are available from the corresponding author upon reasonable request.


\end{document}